\documentclass[twocolumn,showpacs,preprintnumbers,amsmath,amssymb]{revtex4}

\usepackage{graphicx}
\usepackage{dcolumn}
\usepackage{bm}

\renewcommand{\v}[1]{{\bf #1}}

\newcommand{\w}{{\omega}}

\newcommand{\ba}{\begin{eqnarray}}
\newcommand{\ea}{\end{eqnarray}}
\newcommand{\nn}{\nonumber\\}
\newcommand{\Eq}[1]{Eq.~(\ref{#1})}

\newcommand{\sgn}{{\rm sgn}}

\newcommand{\ra}{\rightarrow}

\begin{document}

\preprint{RevTex4}

\title{Quasiparticle Scattering Interference in High Temperature
Superconductors}
\author{Qiang-Hua Wang$^{a,b}$, and Dung-Hai Lee$^{a,c}$}
\affiliation{${(a)}$Department of Physics,University of California
at Berkeley, Berkeley, CA 94720, USA} \affiliation{${(b)}$ National
Laboratory of Solid State Microstructures,\\
Institute for Solid State Physics, Nanjing University, Nanjing
210093, China}\affiliation{${(c)}$ Center for Advanced Study,
Tsinghua University, Beijing 100084, China}


\date{6 May, 2002; Revised August 8}

\begin{abstract}
We propose that the energy-dependent spatial modulation of the
local density of states seen by Hoffman {\it et al}~\cite{hoff2}
is due to the scattering interference of quasiparticles. In this
paper we present the general theoretical basis for such an
interpretation and lay out the underlying assumptions. As an
example, we perform exact T-matrix calculation for the scattering
due to a single impurity. The results of this calculation is used
to check the assumptions, and demonstrate that quasiparticle
scattering interference can indeed produce patterns similar to
those observed in Ref.~\cite{hoff2}.
\end{abstract}

\maketitle

Lately the possibility that a competing charge density wave order
exists in $Bi_2Sr_2CaCu_2O_{8+\delta}$ has received a lot of
attention in scanning tunnelling microscopy (STM) studies. For
example, by applying a magnetic field Hoffman {\it et al} observed
a four-lattice-constant modulation of the local density of states
(LDOS) along the $(0,\pm 1)$ and $(\pm 1,0)$ crystalline
directions in regions surrounding the cores of superconducting
vortices\cite{hoff1}. Quite recently Howald {\it et al} reported
similar LDOS modulation in {\it zero} magnetic field for a
relatively wide range of bias voltage\cite{howald}. These results
have stimulated considerable interests in the high-$T_c$
community\cite{cdws}. Recently Hoffman {\it et al} re-examined the
zero-field LDOS modulations using high resolution
Fourier-transform-scanning tunnelling-spectroscopy\cite{hoff2}.
They found a very important result - the period of the LDOS
modulation changes with bias voltage. One of us (DHL) proposed
that this phenomenon can be explained in terms of the
quasiparticle scattering interferences (QSI).

Scattering interference is an elementary concept in quantum
mechanics. In a nut shell, an incoming wave is scattered into
outgoing wave by some kind of perturbation; the interference
between the two partial waves gives rise to a spatial modulation
of the amplitude of the total wave. LDOS modulation due to the
scattering interference has been long observed for normal
metals\cite{normal}. In the context of high-$T_c$, it has been
suggested that the LDOS modulation due to quasiparticle
interference bears interesting information on the pairing symmetry
of Cooper pairs\cite{byer}.

The purpose of the present paper is to lay down the theoretical
basis for the QSI interpretation of Hoffman {\it et al}'s data. In
particular we spell out the key assumption which allows the
deduction of the normal state Fermi surface and the
superconducting gap dispersion from the observed LDOS
modulations\cite{hoff2,kyle}. As a sanity check, we perform exact
(numerical) scattering calculation for the case of a single
impurity. The results are used to demonstrate the validity of our
assumption and that QSI can indeed produce LDOS modulation
patterns similar to those observed by STM.

Now we briefly summarize the findings of Ref.\cite{hoff2}. By
examining the magnitude of the Fourier transformed LDOS Hoffman
{\it et al} found two discernible groups of modulation wave
vectors. The first is in the $(0,\pm 1)$ and $(\pm 1,0)$
directions and the second is in the $(\pm 1,\pm 1)$ directions.
The wave vectors $\v q$ of the first/second group disperse with
bias voltage $V$ in such a way that $|\v q|$ decreases/increases
when $|V|$ increases. In addition, for fixed $V$,  the $|\v q|$ in
the first/second group decreases/increases with doping. In
Ref.\cite{hoff2} this two group of wavevectors are identified as
the momentum transfer when a quasiparticle is elastically
scattered across the normal state Fermi surface as indicated by
the numbered arrows $1$ and $2$ in Fig.1. Based on this
interpretation Hoffman {\it et al} were able to deduce the Fermi
surface location and superconducting gap function, and found
fairly good agreement with the results of angle-resolved
photoemission experiment\cite{ding}.

Formally the idea used in Ref.\cite{hoff2,kyle} can be precisely
stated in the following Greens function formalism. Let $G_0(\v
k,\w)$ be the ($2\times 2$) single-particle Greens function in the
superconducting state in the absence of scattering, and $T(\v k+\v
q,\v k;\w)$ be the $2\times 2$ scattering matrix for the $\v k \ra
\v k +\v q$ transition. The Fourier transform of the local density
of states is given by \ba n(\v q,\w)&&=n_0(\v q,\w)- \frac{1}{2\pi
i}[A_{11}(\v q,\w)+ A_{22}(\v q,-\w)\nn&&-A_{11}^*(-\v
q,\w)-A_{22}^*(-\v q,-\w) ]\nn A(\v
q,\w)&&=\int{{d^2k}\over{(2\pi)^2}}G_0(\v k+\v q ,\w)T(\v k+\v
q,\v k;\w)G_0(\v k,\w).\nn&&\label{tm}\ea In the above the
subscript ''0'' implies the absence of scattering. Given \Eq{tm}
if one makes the assumption (the ``on-shell'' assumption)  that
the integral in $A(\v q,\w)$ is dominated by those $\v k$'s that
satisfy the simultaneous pole equations for both $G_0$'s, one
obtains the result needed for the QSI interpretation of
Ref.\cite{hoff2,kyle}. In particular one concludes that at a given
bias $eV=\w$, $|n(\v q,\w)|$ is the largest when there is a large
joint density of states associated with scattering wavevector $\v
q$.

The normal state Fermi surface of Bi-2212 consists of four
hole-like segments.\cite{arpes} To get a feel of the joint density
of states let us first concentrate on the curves of constant
energy (CCE) for the quasiparticles. (Since experimentally LDOS
modulation is only observed for energies lower than $\sim 30$ meV,
we shall concentrate on that energy range here.) Around each of
the four gap nodes the CCE evolves from a single point at zero
energy, to banana-shaped closed contours at higher energies. The
size of the banana increases with the energy until its tips touch
the Brillouin zone. When that happen the CCE changes from closed
to open contours. This is schematically shown in Fig.1, where
different CCE's represent different quasiparticle energies. Due to
the large difference between $v_F$ and $v_{\Delta}$
($v_F/v_{\Delta}\approx 20$ at the gap nodes, $v_F=$ the normal
state Fermi velocity, and $v_{\Delta}=$ the derivative of the
superconducting gap along the direction tangential to the Fermi
surface) the CCE moves with energy the fastest at the tips of the
banana. As a result we expect those $\v q$'s connecting the tips
to have the largest joint density of states. This argument leads
to the prediction that the most prominent LDOS modulation wave
vectors should be the set of vectors connecting pairs of banana
tips\cite{kyle}.

\begin{figure}
\includegraphics[width=8.5cm,height=4.75cm]{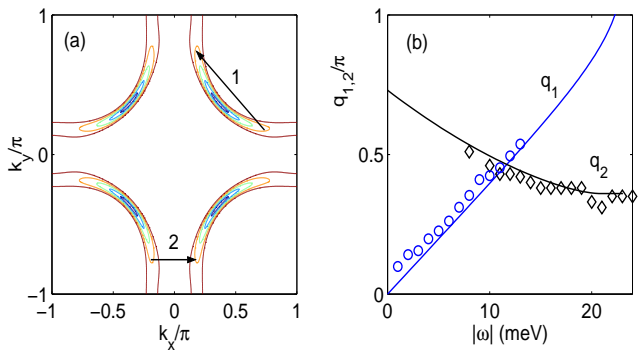}
\caption{(a) Schematic plot of banana-shaped equal-energy
contours. Contours at higher energies are not shown. The tips of
the bananas traces the normal state Fermi surface. The arrows
indicate important elastic scattering processes discussed in the
text. (b) The energy dependence of the modulation wave vector
$q_{1,2}$ associated with the scattering processes 1 and 2
indicated in (a). The solid lines are naive expectations, and the
symbols are extracted from such pictures as plot in Fig.2. A
symbol is put down for a particular energy only when there is a
discernable peak in the wave vector space. It is interesting to
note that only in a narrow energy range peaks in both direction
coexist.}
\end{figure}

As we pointed out earlier, the above expectations are based on the
on-shell approximation. The result of an exact scattering
calculation using \Eq{tm} can in principle deviate from the above
expectations due to the following complications. 1) The
``off-shell'' contributions: i.e. those $\v k$'s that do not
fulfill the simultaneous pole equations of the two $G_0$'s in
\Eq{tm}. These off-shell contributions act to blur the modulation
wave vectors predicted above. 2) The effects of scattering matrix
element. Even when the on-shell approximation works well, the
intensity of the actual LDOS modulation can depend on the actual
scattering matrix element, which in turn depends on the coherence
factor. For example when the scattering is due to a single scalar
($V_s$) and/or magnetic ($V_m$) impurity, the $\v k\ra\v k'$
scattering matrix element is proportional to $
(V_m+V_s)u_ku_{k'}+(V_m-V_s)v_kv_{k'}$, with
$u_k=\pm\sgn(\Delta_k)\sqrt{(1\pm\epsilon_k/E_k)/2}$,$v_k=\sqrt{(1\mp\epsilon_k/E_k)/2}.$
In the above the upper/lower sign applies for positive/negative
energy quasiparticle states, $\epsilon_k$ is the normal state
dispersion, $\Delta_k$ is the gap function, and
$E_k=\sqrt{\epsilon_k^2+\Delta_k^2}$. 
For d-wave pairing $\Delta_k$ changes sign in some of the
scattering  processes and not in others. For example in Fig.1
$\Delta_k$ changes sign in the process labelled by the numbered
arrow 1, while maintains the same sign in the process labelled by
the numbered arrow 2. Therefore depending on whether the
scattering process is caused by scalar or magnetic impurity
processes 2 or 1 will be relatively suppressed.

In the following we perform an exact calculation where the
scattering is caused by a single impurity. The purpose of this
calculation are two folds: 1) through an exact calculation we can
check the on shell assumption that is crucial to interpretation of
Hoffman {\it et al}'s data. 2) Through the exact calculation we
can find out whether the effect of scattering matrix element can
make the scattering interference patterns different from those
expected from the joint density of state argument. This type of
calculation (i.e. an impurity in an d-wave superconductor) is
certainly not new\cite{old}. However as far as we know it is the
first that demonstrate QSI can indeed give rise to LDOS modulation
patterns observed by STM\cite{hoff2,kyle}. Here is the summary of
our results: 1) we find that among the possible wavevectors the
most pronounced LDOS modulation are associated scattering
processes 1 and 2 (and their symmetry equivalence) in Fig. 1. The
Fourier transforms of the LDOS are given in Fig.2. A comparison
between the $\v q-\w$ dispersion expected from the joint density
of states argument and that extracted from Fourier transforming
the exact LDOS modulation is given in Fig.1(b).

For the quasiparticle Hamiltonian, we use a model provided by
Norman {\it et al}\cite{norman}:
\begin{equation} H=\sum_{k\sigma}\epsilon_k
C_{k\sigma}^{\dagger}C_{k\sigma}-\sum_k(\Delta_kC_{k\uparrow}C_{-k\downarrow}
+{\rm h.c.}). \end{equation} In the above $\epsilon_k=\sum_{n=0}^5
t_n\chi_n(k)$ and $\Delta_k=\Delta_0(\cos k_x-\cos k_y)/2$, where
$t_{0-5}=0.1305$, $-0.5951$, $0.1636$, $-0.0519$, $-0.1117$,
$0.0510$ ($eV$), and $\Delta_0=0.025eV$. Moreover
$\chi_{0-5}(k)=1$, $(\cos k_x+\cos k_y)/2$, $\cos k_x\cos k_y$,
$(\cos 2k_x+\cos 2k_y)/2$, $(\cos 2k_x\cos k_y+\cos 2k_y\cos
k_x)/2$, and $\cos 2k_x\cos 2k_y$. The $G_0(\v k,\w)$ and $T(\v
k+\v q, \v k;\w)$ [$=T(\w)$ for a single impurity] in
Eq.(\ref{tm}) are given by, respectively, \ba &&G_0^{-1}(\v
k,\w)=(\w+i\delta) I -\epsilon_\v k\sigma_3-\Delta_\v k\sigma_1\nn
&&T^{-1}(\w)=(V_s\sigma_3+V_m I)^{-1}-\int{{d^2k}\over{(2\pi)^2}}
G_0(\v k,\w),\label{Eq:Tmatrix} \ea where $\sigma_i$'s are the
Pauli matrices.

\begin{figure}
\includegraphics[width=8.5cm,height=8cm]{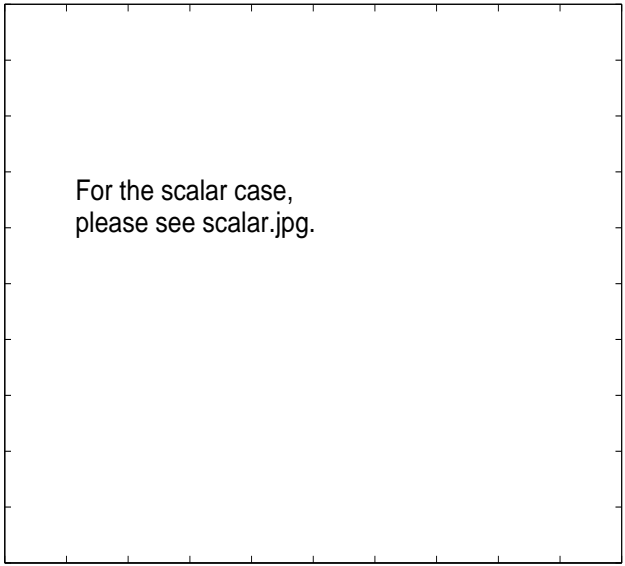}
\caption{Fourier amplitudes of the LDOS maps at the given energies
as a function of momenta in the first Brillouin zone for
$(V_s,V_m)=(100meV,0)$. The numbers on the sub-graphs show the
energy in units of meV. The hot/cold color denotes higher/lower
strength. }
\end{figure}

\begin{figure}
\includegraphics[width=8.5cm,height=8cm]{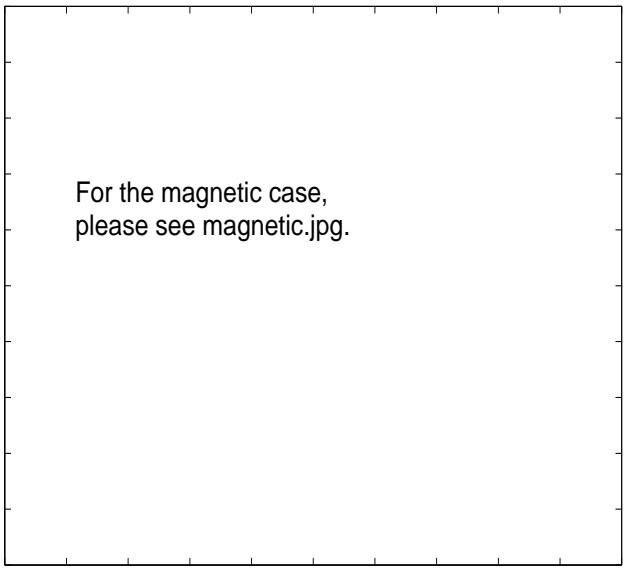}
\caption{The same as Fig.2 for $(V_s,V_m)=(0,100meV)$. }
\end{figure}

In order to accurately take into account the contribution to $G_0$
from the thin bananas in Fig.1(a), a momentum resolution $\delta
k\ll \pi v_{\Delta}/v_F$ must be achieved. This in turn means that
a large real space lattice has to be used for the actual
calculation. For the results reported below a $400\times 400$
lattice is used. In this calculation we did not take the
self-consistent suppression of the pairing amplitude near the
impurity into account. However we do not expect this omission to
cause any significant change because the impurity scattering
strength we used is much larger than the maximal pairing
gap.\cite{gapimp}

In the following two kinds of impurities are studied: (a)
non-magnetic impurity ($V_s=100meV$ and $V_m=0$) and (b) magnetic
impurity ($V_s=0$ and $V_m=100meV$). By using two extreme types of
scatterer we can study the effect of matrix element on the QSI
pattern. The impurity strength we choose is intermediate in the
sense that we do not see quasiparticle resonance states in the
LDOS.\cite{balatsky} It is interesting that although this type of
impurity does not show up in the zero bias resonance image, they
do have important effect on the LDOS modulation.

Fig.2 shows the Fourier amplitude of the LDOS maps at  $\w=-24
\rightarrow -3$ meV and $\w=3\rightarrow 24$ meV with 3 meV
interval for a nonmagnetic. In these figures the intensity at $\v
q =0$ is subtracted so that weaker features at other wave vectors
can show up. In constructing Figure 2 the LDOS of the central
$51\times 51$ plaquette in the $400\times 400$ lattice are Fourier
transformed. It is interesting to note that in real experiment a
large field of view is also employed in the Fourier transform in
order to achieve high momentum resolution.

For the negative energies in Fig.2 the Fourier peaks in the $(\pm
1,\pm 1)$ direction is clearly visible between $-3$ to $-12$ meV.
Moreover as the binding energy increases the peaks moves away from
the origin. For binding energy 18 meV and above the Fourier peaks
in the $(0,\pm 1)$ and $(\pm 1,0)$ appear. Contrary to the
diagonal direction spots these peaks move only slightly as the
energy varies. Careful analysis of such movement indicates that
these peaks move toward the origin as the binding energy
increases. The energy dependence of the these LDOS Fourier peaks
is shown in Figure 1(b) as symbols. The difference between the
naive expectation (solid lines) and the results of T-matrix
calculation (symbols) is very small. This comparison clarifies the
extent by which the qualitative picture of joint density of states
works. In addition to the above peaks there are other weaker
features in Fig.2. These features can either be identified as the
higher harmonics of the diagonal and vertial/horizontal spots or
can be attributed to the other scattering processes not shown in
Fig.1\cite{kyle}. Upon the reversal of the bias voltage we find
the diagonal Fourier peaks become much weaker. On the contrary the
$(0,\pm 1)$ and $(\pm 1,0)$ direction peaks seem to be insensitive
to the bias reversal\cite{note}.

The quasiparticle interference patterns for magnetic impurity in
Fig.3 are quite different from those shown in Fig.2. For example
the amplitude of the $(\pm 1,\pm 1)$ direction modulation for
$|\w|\leq 12$ meV is much weaker. The fact that the $(\pm 1,\pm
1)$ direction peaks are weaker than the $(\pm 1,0)$ and $(0,\pm
1)$ direction peaks is consistent with the coherence factor effect
discussed previously.

In addition to the above differences there are similarities
between Figs.2 and 3. For example at $|\w|=21,24$ meV, the
patterns in these figures become rather similar. In the same
energy range the asymmetry between positive and negative bias
becomes small. In this energy range the horizontal component of
the scattering wavevector associated with process 2 cease to
change with energy. As the result the modulation wave-vectors in
the $(\pm 1,0)$ and $(0,\pm 1)$ directions hardly change.

Upon closing a few remarks are in order. 1) The present result
demonstrates that QSI due to scattering by scalar (non-unitary)
impurities can produce LDOS modulation similar to that observed in
recent STM experiment\cite{hoff2,kyle}. This effect is due to
quantum interference, and does not require the presence of a
charge density wave order. 2) Whether the phenomenology of a very
weak/fluctuating charge density wave order is consistent with the
experimental observations\cite{hoff2,kyle} should be further
investigated\cite{howald,steve}. 3) The results of the present
paper is obtained assuming sharply defined quasiparticles in the
superconducting state\cite{zxs}. However our general formulation
(\Eq{tm}) allows us to treat not so well defined quasiparticles as
well - all we have to do is to modify the sharp pole structure of
$G_0$. We find that the $(\pm 1,0),(0,\pm 1)$ direction modulation
is robust against quasiparticle life time broadening whereas the
$(\pm 1,\pm 1)$ direction modulation is not. 4) In the present
paper we use a single impurity as a representative of elastic
quasiparticle scatterer. However we believe that the QSI idea
should work under wider circumstances. 5) It is not clear to us
what is the main source of quasiparticle scattering in BSCCO. For
example, recently Podolsky {\it et al} argued that scattering by a
disordered charge density wave could give rise to similar energy
dependent $(\pm 1, 0)$ or $(0,\pm 1)$ modulations as
well\cite{pod}. 6) Finally, what is the relation between the
relatively weak quasiparticle interference observed at zero
magnetic field\cite{hoff2,kyle} and the relatively strong
checkerboard LDOS modulation near superconducting
vortices\cite{hoff1} is currently unclear.

{\bf{Acknowledgments}}: This work is supported by DMR 99-71503.
QHW is also supported by the Natural Science Foundation of China,
and the Grant for State Key Program of China grant No.
G1998061407. We thank J.C. Davis's group at Berkeley for sharing
their unpublished data with us. We are also grateful for the
helpful discussions with J.C. Davis, M. E. Flatte, Jung Hoon Han,
J.E. Hoffman, S.A. Kivelson, K. McElroy, J. Orenstein, Z. X. Shen,
R. W. Simmonds, and C. M. Varma.


\begin{references}
\bibitem{gapimp} In the presence of order parameter
suppression there is an additional anomalous scattering potential
(due to the change in the gap parameter) around the location of
the impurity. However the strength of this anomalous scattering
potential is bounded above by the maximal pairing gap in the bulk.
Since the scalar potential is much stronger its effects dominate.
We have done extensive calculations where there are both scalar
and anomalous scattering potentials. The above statement has been
checked carefully.
\bibitem{note} We have investigated the asymmetry of
the $(\pm 1,\pm 1)$ direction Fourier peaks upon bias reversal
further. We find such asymmetry depends on the detailed
quasiparticle dispersion. For example by setting all the $t_n$
except $n=1$ to zero we find the asymmetry disappears. Thus we
conclude that we We have investigated the issue of plus/minus bias
voltage asymmetry of the $(\pm 1,\pm 1)$ direction Fourier peaks
further. We find the detailed quasiparticle dispersion has
important effects on this asymmetry. For example by setting all
the $t_n$ except $n=1$ to zero we find the asymmetry disappears.
Therefore we conclude that we can not make a definitive prediction
of the bias asymmetry.
\end{references}
\end{document}